\documentclass[12pt,a4,portrait]{article}
\usepackage{latexsym}
\usepackage{hyperref}
\usepackage{graphicx}

\newcommand{\be}{\begin{equation}}
\newcommand{\ee}{\end{equation}}
\newcommand{\beq}{\begin{eqnarray}}
\newcommand{\eeq}{\end{eqnarray}}
\newcommand{\bed}{\begin{displaymath}}
\newcommand{\eed}{\end{displaymath}}
\newcommand{\bc}{\begin{center}}
\newcommand{\ec}{\end{center}}
\newcommand{\bi}{\begin{itemize}}
\newcommand{\ei}{\end{itemize}}
\newcommand{\bn}{\begin{enumerate}}
\newcommand{\en}{\end{enumerate}}

\newcommand{\rw}{ {\rm w} }
\newcommand{\dd}{ {\rm d} }
\newcommand{\pmp}{ \matrix{ + \vspace{-2mm}\cr \mbox{\scriptsize(} \! - \! \mbox{\scriptsize)}} }
\newcommand{\mpp}{ \matrix{ - \vspace{-2mm}\cr \mbox{\scriptsize(} \! + \! \mbox{\scriptsize)}} }
\newcommand{\mxp}{ \matrix{ - \vspace{-2mm}\cr \mbox{\scriptsize\phantom(} \! \phantom+ \! \mbox{\scriptsize\phantom)} } }
\newcommand{\xpp}{ \matrix{ \phantom- \vspace{-2mm}\cr \mbox{\scriptsize(} \! + \! \mbox{\scriptsize)}} }

\begin{document}
\baselineskip 1.2\baselineskip
\thispagestyle{empty}

{\phantom .}

\begin{center}


{\Large\bf{Scalar-tensor cosmologies: \linebreak
fixed points of the Jordan frame scalar field}}

\vspace{2.0 cm}

Laur J\"arv,$^{1 a}$  Piret Kuusk,$^{2 a}$ and Margus Saal$^{3  b}$

\vspace{0.3cm}
{\it {\phantom x}$^{a}$ Institute of Physics, University of Tartu,
Riia 142, Tartu 51014, Estonia \linebreak
{\phantom x}$^{b}$ Tartu Observatory, T\~oravere 61602, Estonia}

\end{center}

\vspace{1,5cm}

\begin{abstract}
We study the evolution of homogeneous and isotropic, flat cosmological models within the general
scalar-tensor theory of gravity with arbitrary coupling function and potential. After introducing the
limit of general relativity we describe the details of the phase space geometry. Using the methods
of dynamical systems for the decoupled equation of the Jordan frame scalar field we find the fixed
points of flows in two cases: potential domination and matter domination.
We present the conditions on the mathematical form of the coupling function and potential which 
determine the nature of the fixed points (attractor or other).
There are two types of fixed points, both are characterized by cosmological evolution mimicking 
general relativity, but only one of the types is compatible with the Solar System PPN constraints.
The phase space structure should also carry over to the Einstein frame as long as the transformation 
between the frames is regular which however is not the case for the latter (PPN compatible) fixed point. 
\end{abstract}

\vspace{1cm}

\vspace{3 cm}
$^1$ Electronic address: laur@fi.tartu.ee

$^2$ Electronic address: piret@fi.tartu.ee

$^3$ Electronic address: margus@fi.tartu.ee

\newpage
\section{Introduction}

Scalar-tensor theories (STT) of gravitation \cite{will, fm, faraoni} 
have emerged in different contexts of theoretical physics, e.g. 
in Kaluza-Klein type unifications, supergravity, and
low energy approximations of string theories.
The scalar-tensor action functional has been also used
in topology as a mathematical tool for solving the geometrization conjecture of
3-manifolds via the Ricci flow \cite{anderson}. 

In cosmology STT has been invoked to model the accelerated expansion of
 inflation  \cite{inflation} and dark energy \cite{darkenergy} (for some
recent papers see e.g. Ref. \cite{darkenergy_recent}).
However, observations in the Solar System tend to indicate that in the 
intermediate-range distances the present Universe around us is 
successfully described by 
the Einstein tensorial gravity alone \cite{tests}. 
This means that only such models of scalar-tensor gravity are viable which 
in their late time cosmological evolution
imply local observational consequences very close to those of Einstein's 
general relativity (GR) \cite{tumty}.

The importance of the convergence of the STT solutions to those of 
GR 
at late times, 
guaranteeing 
agreement with the present post-Newtonian observations was recognized
by Bekenstein and Meisels \cite{bek} already long ago. 
Damour and Nordtvedt \cite{dn0, dn} 
argued that for a wide class of homogeneous and isotropic 
scalar-tensor cosmological models 
there exists an attractor mechanism taking them to the limit of  
the corresponding general relativistic models.
In subsequent papers the issue has been pursued by various approaches
\cite{asymptotics, bp, san, bs}.

The methods of dynamical systems provide natural 
tools to analyze the problem.
Most previous studies which have considered STT cosmology 
as a dynamical system have focused upon examples with specific coupling
functions, see Refs.  
\cite{dynsys_einstein, mn, carloni_2008, meie3} for works based on the Einstein frame and
Refs. \cite{dynsys_jordan, gunzing_2001, abramo_2003, meie2, meie3} for the Jordan frame.
In the Jordan frame the main properties of 
the general phase space geometry 
were outlined
by Faraoni \cite{faraoni1}, while Refs.  \cite{burd_coley, faraoni3}
consider general Jordan frame dynamics 
with particular attention to the de Sitter fixed point.

Although the equations of motion in the Einstein frame may have a simpler form,
the cosmological observations are easier to interpret
in the Jordan frame where matter is minimally coupled.
In general there is a one-to-one 
correspondence between the phase spaces in the
two conformal 
frames \cite{dicke, weinberg, faraoni, flanagan, faraoni2},
provided that the transformations from one frame to another are regular.
However, 
as it was pointed out in Ref. \cite{meie3}, the differential equation
relating the 
scalar fields in these two frames becomes singular at the 
limit of general relativity.
Therefore utmost care must be exercised when addressing the question
whether general relativity is an attractor, as the answer may depend 
on the frame chosen.

In this paper we take the Jordan frame and consider general 
scalar-tensor theories which 
contain two functional degrees of freedom, the coupling function 
$\omega (\Psi)$ and the scalar potential $V(\Psi)$. 
We perform the dynamical systems 
analysis for the flat Friedmann-Lema\^{\i}tre-Robertson-Walker (FLRW)
backgrounds with ideal barotropic fluid matter.
Our strategy is to find the fixed points for
the scalar field dynamics and compare these 
with the conditions of the limit of general relativity in the Solar System,
as established by the parameterized post-Newtonian (PPN) formalism.
Therefore, if the functional forms of $\omega (\Psi)$ and $V(\Psi)$
are specified from some considerations (e.g. the compactification manifold), 
our results allow to determine the fixed points along with their type and thus
immediately decide whether general relativity is an attractor, 
i.e. whether the model at hand is viable or not.

The plan of the paper is the following. 
In section 2 we 
write the field equations 
as a dynamical system and find
the conditions which reduce these equations to the FLRW equations
of general relativity (possibly with a cosmological constant),
concluding that these conditions are marginally stricter 
than the standard PPN conditions.
In section 3 we describe the general phase space
and its physical subspace determined by the Friedmann constraint equation.    
In section 4 we consider the case of potential domination with
(effectively) vanishing matter density,
find the fixed points and determine their properties.
For investigating the case of matter domination (with vanishing
scalar potential) in section 5, we follow Refs. \cite{dn} and \cite{san} 
to introduce a new time parameter which allows us
derive a decoupled ``master'' equation for the scalar field
and again perform the fixed point analysis.  
In section 5 we briefly discuss comparison to
some previous works and 
end with a speculation about the possible relevance of our results for
the moduli stabilization problem in string theory. 
Section 6 gives a summary.

\section{Scalar-tensor cosmology as a dynamical system and the limit of general relativity}

We consider  a general scalar-tensor theory
in the Jordan frame given by the action functional
    \beq \label{jf4da}
S  = \frac{1}{2 \kappa^2} \int d^4 x \sqrt{-g}
        	        \left[ \Psi R(g) - \frac{\omega (\Psi ) }{\Psi}
        		\nabla^{\rho}\Psi \nabla_{\rho}\Psi
                  - 2 \kappa^2 V(\Psi)  \right]
                   + S_{m}(g_{\mu\nu}, \chi_m) \,.
\eeq 
Here $\omega(\Psi)$ is a coupling function and $V(\Psi)$ is a
scalar potential, $\nabla_{\mu}$ 
denotes the covariant derivative with respect to the metric 
$g_{\mu\nu}$, $\kappa^2=8 \pi G_N$, and 
$S_{m}$ is the matter part of the action 
as all other fields are included in $\chi_m$.
In order to keep the effective Newtonian gravitational constant positive
\cite{nordtvedt70,bp} we assume that $0 < \Psi < \infty$.

The field equations for the flat ($k=0$) 
FLRW line element and perfect barotropic fluid 
matter, $p=\rw \rho$, read 
\beq 
\label{00}
H^2 &=& 
- H \frac{\dot \Psi}{\Psi} 
+ \frac{1}{6} \frac{\dot \Psi^2}{\Psi^2} \ \omega(\Psi) 
+ \frac{\kappa^2}{3} \frac{ \rho}{\Psi} 
+ \frac{\kappa^2}{3} \frac{V(\Psi)}{\Psi} \,, 
\\ \nonumber \\
\label{mn}
2 \dot{ H} + 3 H^2 &=& 
- 2 H \frac{\dot{\Psi}}{\Psi} 
- \frac{1}{2} \frac{\dot{\Psi}^2}{\Psi^2} \ \omega(\Psi) 
- \frac{\ddot{\Psi}}{\Psi} 
- \frac{\kappa^2}{\Psi} \rw \rho 
+ \frac{\kappa^2}{\Psi} \ V(\Psi) \,, 
\\ \nonumber \\
\label{deq}
\ddot \Psi &= & 
- 3H \dot \Psi 
- \frac{1}{2\omega(\Psi) + 3} \ \frac{d \omega(\Psi)}{d \Psi} \  \dot {\Psi}^2 
+ \frac{\kappa^2}{2 \omega(\Psi) + 3} \ (1-3 \rw) \ \rho  \nonumber \\ 
& &\qquad + \frac{2 \kappa^2}{2 \omega(\Psi) + 3} \ \left[ 2V(\Psi) - \Psi \ 
\frac{d V(\Psi)}{d \Psi}\right] \, ,
\eeq 
$H \equiv \dot{a} / a$. 
The matter conservation law is the usual 
\beq \label{matter_conservation}
\dot{\rho} + 3 H \ (\rw+1) \ \rho = 0 \,
\eeq
and we assume $\rho \geq 0$.
Eqs. (\ref{00})--(\ref{matter_conservation}) 
are too cumbersome to be solved analytically, 
but useful information about the general characteristics of solutions
can be obtained by rewriting (\ref{00})--(\ref{matter_conservation}) 
in the form of 
a dynamical system and finding
the fixed points which describe the asymptotic behaviour of solutions.

The phase space of the system
is spanned by four variables $\{ \Psi, \dot{\Psi}, H, \rho \}$. 
Defining $\Psi \equiv x, \dot{\Psi} \equiv y$ the dynamical system corresponding to
equations 
(\ref{00})-(\ref{matter_conservation}) can be written
as follows:
\beq
\label{general_dynsys_x}
\dot{x} &=& y \,, \\ 
\label{general_dynsys_y}
\dot{y} &=& -\frac{1}{2\omega(x)+3} \left[ 
\frac{d\omega(x)}{dx} y^2 -\kappa^2 \left( 1-3\rw \right)\rho
+2 \kappa^2 \left( \frac{dV(x)}{dx}x-2V(x) \right)
 \right] -3Hy \,, \\
\label{general_dynsys_H}
\dot{H} &=& \frac{1}{2x(2\omega(x)+3)} \left[ 
\frac{d\omega(x)}{dx} y^2 -\kappa^2 \left( 1-3\rw \right)\rho
+2 \kappa^2 \left( \frac{dV(x)}{dx}x -2V(x) \right)
 \right] \nonumber \\ 
&& \quad -\frac{1}{2x} \left[ 6H^2x + 2Hy - \kappa^2 (1-\rw)\rho -2\kappa^2 V(x)
\right] \,, \\
\dot{\rho} &=& -3 H (1+\rw) \rho \,.
\label{general_dynsys_rho}
\eeq

Based on these equations we may 
make a couple of quick qualitative observations about 
some general features of the solutions. 
For example, the limit $\Psi \rightarrow 0$ in general implies 
$|\dot{H}| \rightarrow \infty$,
hence the solutions can not safely pass from positive to negative values of 
$\Psi$
(from ``attractive'' to ``repulsive'' gravity), 
but hit a space-time singularity as the curvature invariants diverge.
Similarly, the limit $2\omega + 3 \rightarrow 0$ implies 
$|\dot{H}| \rightarrow \infty$ with the same conclusion that 
passing through $\omega(\Psi)=-\frac{3}{2}$ 
(corresponding to the change of the sign of the scalar field kinetic
term in the Einstein frame action)
would entail a space-time singularity and is impossible. 
(Let us remark here, that these observations are quite general and do not 
preclude specially fine-tuned solutions in some fine-tuned models which may 
remain regular while crossing these points, like in Ref. \cite{gunzing_2001}.)

The limit $\frac{1}{2\omega + 3} \rightarrow 0$ deserves a more 
closer examination. Let us define $x_{\star}$ by
\be \label{x_star}
\frac{1}{2\omega(x_{\star}) + 3} =0 \,.
\ee
Expressing $H$ from the Friedmann constraint (\ref{00}),
\be
\label{H_constraint}
H = -\frac{y}{2 x} \pmp
\sqrt{\left( 2\omega(x)+3 \right) \frac{y^2}{12x^2} 
+ \frac{\kappa^2 (\rho + V(x))}{3x} } \, ,
\ee
makes clear that $|H|$ diverges as $x \to x_{\star}$, 
unless also $y \to 0$ at the same time. 
What happens in the latter case is
determined by the first term under the square root above. We can
compute its limit by Taylor expanding
\beq
\lim_{y \to 0 \atop x \to x_{\star}} (2\omega(x)+3)y^2 &=&
\lim_{\Delta y \to 0 \atop \Delta x \to 0} 
\frac{y^2|_{y=0} + 2 y|_{y=0} \Delta y + (\Delta y)^2}
{\frac{1}{2\omega(x)+3}\Big|_{x=x_{\star}} 
+ \frac{ \dd}{\dd x}\left(\frac{1}{2\omega(x)+3}\right)\Big|_{x=x_{\star}}\Delta x
+ \frac{1}{2} \frac{\dd^2}{\dd x^2}\left(\frac{1}{2\omega(x)+3}\right)\Big|_{x=x_{\star}} (\Delta x)^2 + \dots  } \nonumber \\
&=& \lim_{r \to 0} \frac{r^2 \sin^2\theta}
{\frac{\dd}{\dd x}\left(\frac{1}{2\omega(x)+3}\right)\Big|_{x=x_{\star}} r \cos\theta 
+ \frac{1}{2} \frac{d^2}{dx^2}\left(\frac{1}{2\omega(x)+3}\right)\Big|_{x=x_{\star}} r^2 \cos^2\theta + \dots} \nonumber \\
\vspace{15mm}
&&
\left \{ \begin{array}{ll}
= 0 \, , & $if$ \quad \frac{\dd}{\dd x}\left(\frac{1}{2\omega(x)+3}\right)\Big|_{x=x_{\star}} \neq 0 \,, \vspace{2mm}\\ 
\sim \tan^2 \theta \, , &  $if$ \quad  \frac{\dd}{\dd x}\left(\frac{1}{2\omega(x)+3}\right)\Big|_{x=x_{\star}}=0 \; , \quad  \frac{\dd^2}{\dd x^2}\left(\frac{1}{2\omega(x)+3}\right)\Big|_{x=x_{\star}} \neq 0 \,, \vspace{2mm}\\
= \infty \, , & $if$ \quad  \frac{\dd}{\dd x}\left(\frac{1}{2\omega(x)+3}\right)\Big|_{x=x_{\star}}= 0 \;, \quad \frac{\dd^2}{\dd x^2}\left(\frac{1}{2\omega(x)+3}\right)\Big|_{x=x_{\star}} = 0 \,,
\end{array}
\right. 
\label{taylor_expansion}
\eeq
where $\Delta x = r \cos\theta$, $\Delta y = r \sin\theta$ was taken.
(We have neglected the unphysical direction $|\theta| = \frac{\pi}{2}$
 that corresponds
to approaching the point $(x=x_{\star},y=0)$ 
along the line $x=x_{\star}$ where
 $|H|$ is divergent.)
So, in the limit $x \to x_{\star}, y \to 0$ 
the value of $H$ is determined by the lowest non-zero derivative of
$\frac{1}{2\omega(x) + 3}$. 
If the both the first and second derivative vanish, 
then $|H|$ diverges implying
a spacetime singularity.
If the first derivative vanishes but the second derivative is not zero, 
\be
 \frac{\dd}{\dd x}\left(\frac{1}{2\omega(x)+3}\right)\Big|_{x=x_{\star}}=0 \; , \qquad  \frac{\dd^2}{\dd x^2}\left(\frac{1}{2\omega(x)+3}\right)\Big|_{x=x_{\star}} \neq 0 \,,
\ee
then $H$ is finite but (possibly) different for each solution as it 
depends on the angle of approach $\theta$, while 
the Friedmann equation in this case acquires an extra term 
 when compared to general relativity.
If the first derivative is not zero,
\be \label{first_derivative_nonzero}
 \frac{\dd}{\dd x}\left(\frac{1}{2\omega(x)+3}\right)\Bigg|_{x=x_{\star}}= \frac{1}{(2\omega(x_{\star})+3)^2}\frac{\dd \omega}{\dd x}\Bigg|_{x=x_{\star}} \neq 0 \,,
\ee
then $H$ approaches the value 
$H_{\star}^2 = \frac{\kappa^2}{3 x_{\star}}(\rho+V(x_{\star}))$,
mimicking the Friedmann equation of general relativity
with $8 \pi G = \frac{\kappa^2}{x_{\star}}$ and 
$\Lambda = \frac{\kappa^2}{x_{\star}}V(x_{\star})$. 

To summarize, we have just observed that in the limit 
(a) $\frac{1}{2\omega(x) + 3} \rightarrow 0$,
(b) $y \rightarrow 0$
 the Friedmann constraint (\ref{H_constraint}) tends to the form
of general relativity 
if 
(c) $\frac{1}{(2\omega(x_{\star})+3)^2}\frac{\dd \omega}{\dd x}\Big|_{x=x_{\star}} \neq 0$. 
It must be also emphasized here 
that the process of taking the Taylor expansion (\ref{taylor_expansion})
hinges on the assumption that (d)
 $\frac{1}{2\omega(x)+3}$ is differentiable (derivatives do not diverge)
at $x_{\star}$.
In this context one may also ask 
when the full set of equations 
(\ref{general_dynsys_x})-(\ref{general_dynsys_rho})
attains the form 
of general relativity. It is easy to see that besides 
(a)-(d) one must also impose
\be
\frac{1}{2\omega(x)+3} \frac{\dd \omega}{\dd x} y^2 = 
\frac{1}{(2\omega(x_{\star})+3)^2}\frac{\dd \omega}{\dd x}
\left( (2\omega(x)+3) y^2 \right) \to 0 \,,
\ee
but the latter is automatically satisfied if (c) holds, due to 
(\ref{taylor_expansion}), (\ref{first_derivative_nonzero}).
Therefore we may tentatively call the conditions (a)-(d) 
`the general relativity limit of scalar-tensor flat FLRW 
cosmology'.

It is interesting to compare the cosmological GR limit to the GR limit
obtained from PPN, which
characterizes the slow motion approximation in a centrally symmetric 
gravitational field. Although the mathematical assumptions underlying the 
PPN formalism are clearly different from our
cosmological reasoning above, we may still ask whether the results of both
schemes agree with each other.
In the context of PPN it is well established that 
the solutions of scalar-tensor theory approach those of general relativity
when \cite{nordtvedt70}
\be \label{PPN_condition}
\frac{1}{2\omega(x)+3} \to 0 \,, \qquad
\frac{1}{(2\omega(x)+3)^3} \frac{\dd \omega}{\dd x} \to 0 \,.
\ee
Comparison shows that 
the cosmological conditions (a)-(d) are marginally stricter than the PPN 
condition 
(\ref{PPN_condition}), since (a), (c), (d) imply that (\ref{PPN_condition})
is satisfied, but (\ref{PPN_condition}) does not necessarily guarantee that 
(c) or (d) holds. 

Let us also note that there is also another special case
$x_{\bullet}$, realized at
\be \label{second_GR_limit}
\rho = 0, \qquad
y = 0, \qquad
2V(x_{\bullet}) - x_{\bullet} \ \frac{\dd V(x)}{\dd x}\Big|_{x_{\bullet}} =0 \,,
\ee
when the cosmological equations 
(\ref{general_dynsys_x})-(\ref{general_dynsys_rho})
relax to those of general relativity featuring de Sitter evolution. 
However, as the value of 
$\omega(x_{\bullet})$ is not fixed by the condition (\ref{second_GR_limit}),
this case does not conform with the GR limit of PPN.
Therefore, even when the limits (a)-(d) and 
(\ref{second_GR_limit}) can be cosmologically indistinguishable,  
Solar System observations in the PPN framework can in principle reveal
which of the two is actually realized.
(In this paper when using the phrase `the GR limit of STT' we mean the conditions (a)-(d), 
as these take the STT cosmological equations to those of  general relativity 
and also guarantee that the PPN condition is satisfied.
But note that some authors, e.g. Refs. \cite{mn, carloni_2008}
have not necessarily used the same definition.)

The general relativity limit of STT is
purely given in terms of $x$ and $y$. In the following we extract from the full
dynamical system (\ref{general_dynsys_x})-(\ref{general_dynsys_rho})
an independent subsystem for $\{ x,y \}$, find its fixed points and
check whether the limit of general relativity matches to an attractive fixed
point. 
But before, some general remarks about the full phase space are in order.

\section{Phase space}

In the four phase space dimensions  
$\{ \Psi \equiv x, \dot{\Psi} \equiv y, H, \rho \}$ 
the physical trajectories 
(orbits of solutions) are those
which satisfy the Friedmann constraint (\ref{00}), i.e. which
lie on the 3-surface
\be
\label{Friedmann_surface}
\mathcal{F}: F(x,y,H,\rho)\equiv H^2 + H\frac{y}{x} - \frac{y^2}{6 x^2} \ \omega(x) 
- \frac{\kappa^2 \rho}{3x} -\frac{\kappa^2 V(x)}{3x} =0\,.
\ee
It can be readily checked that 
the trajectories' tangent vector, $T^i=(\dot{x},\dot{y},\dot{H},\dot{\rho})$,
given by (\ref{general_dynsys_x})-(\ref{general_dynsys_rho}), 
is perpendicular to the normal of the Friedmann surface,
\be
 \nabla_i F \cdot T^i \Big|_{\mathcal{F}} = 0,
\ee
and therefore the system automatically obeys the Friedmann constraint, 
as the trajectories on the surface $\mathcal{F}$ never leave it. 

In principle the geometry of the Friedmann 3-surface in the 4-dimensional
phase space is rather complicated to visualize,
but a few general characteristics can still be given.
We may write Eq. (\ref{Friedmann_surface}) as
\be
\frac{\left(H + \frac{y}{2x}\right)^2}{\frac{\kappa^2(\rho+V)}{3x}}
- \frac{y^2}{\frac{4 \kappa^2 x (\rho+V)}{2\omega+3}} = 1 \, ,
\ee
which for fixed $\rho$ and $x$ can be recognized as describing 
familiar conic sections:
1) for $\rho+V>0$, $2\omega+3>0$ a hyperbola 
on the $(H + \frac{y}{2x},y)$ plane, 
2) for $\rho+V>0$, $2\omega+3<0$ an ellipse 
also on the $(H + \frac{y}{2x},y)$ plane,
while
3) for $\rho+V<0$, $2\omega+3>0$ a hyperbola 
on the $(y, H + \frac{y}{2x})$ plane. 
The case 
4) $\rho+V<0$, $2\omega+3<0$ is not realized as 
real solutions are absent.
This result establishes that the intersection of the Friedmann 3-surface
with the (fixed $\rho$, fixed $x$) 2-plane is constituted in either
one piece (ellipse) or two pieces (hyperbola). 
Thus in case 1) the allowed phase space is divided into two separate regions,
the ``upper'' region where $H+\frac{y}{2x}>0$ and 
the ``lower'' region
 where $H+\frac{y}{2x}<0$, and
there is no way the trajectories can travel from one region to another. 
In case 2) these two regions meet along a 2-surface where 
$H+\frac{y}{2x}=0$, and the trajectories can in principle cross
from one region to another. 
In case 3) there are again two separate parts, now characterized by
$y>0$ and $y<0$, respectively.
At first it may be difficult find a direct physical interpretation 
for the quantity $H + \frac{y}{2x}$ that characterizes the 
``upper'' and ``lower'' region in cases 1) and 2), 
but it turns out that this combination is equal to the Hubble parameter in
the Einstein frame, see Eq. (\ref{H_E H_J}), and thus the ``upper''
region corresponds to universes which expand 
in the Einstein frame, 
while 
the ``lower'' region has universes 
which contract in the Einstein frame.

Related information can be also established by another approach.
We may solve the Friedmann constraint for $H$, Eq. (\ref{H_constraint}),
and then the condition for all variables to be real valued 
imposes an inequality
\be
\label{y_allowed_inequality}
 \left( 2\omega(x)+3 \right) \frac{y^2}{12x^2} 
+ \frac{\kappa^2 (\rho + V(x))}{3x} \ge 0 \, .
\ee
In terms of physics 
this inequality can be interpreted as a restriction on the allowed 
values of $y$ (see Table \ref{allowed_regions}). 
There is no restriction in case 1), while
the case 
4) $\rho+V<0$, $2\omega+3<0$ is completely ruled out since no
real solutions compatible with the Friedmann constraint exist in this case.
Similarly, solving the Friedmann constraint for $y$ leads to 
another inequality,
\be
\label{H_allowed_inequality}
\frac{x^2}{\omega^2(x)} \left( (2\omega(x)+3)H^2 - \frac{2 \kappa^2 \omega(x)}{3x} (\rho+V(x)) \right) \ge 0 \, ,
\ee
which can be interpreted as a restriction on the allowed values of $H$
(given also in Table \ref{allowed_regions}). 
Analogously, once $\omega(x)$ and $V(x)$ are specified, 
we may get a third inequality from solving the Friedmann constraint for $x$
as well.
\begin{table}[t] 
\begin{center}
\begin{tabular}{llcccclcl}
&  & \qquad &  & \quad & Allowed range of $\dot \Psi$ && Allowed range of $H$
\vspace{2mm}\\
\hline \\
1a) & $\rho+V \geq 0$ && $0 \leq \omega \leq \infty$ 
&& $0 \leq {\dot\Psi}^2 \leq \infty$ 
&& $\frac{2\kappa^2\omega(\rho+V)}{3\Psi(2\omega+3)} \leq H^2 \leq \infty$
\vspace{2mm}\\
1b) & $\rho+V \geq 0$ && $-\frac{3}{2} \leq \omega \leq 0$ 
&& $0 \leq {\dot\Psi}^2 \leq \infty$ 
&& $0 \leq H^2 \leq \infty$
\vspace{2mm}\\
2) & $\rho+V > 0$ && $-\infty \leq \omega \leq -\frac{3}{2}$ 
&& $0 \leq {\dot\Psi}^2 \leq \frac{4 \kappa^2 (\rho+V) \Psi}{|2 \omega+3|}$
&& $0 \leq H^2 \leq \frac{2\kappa^2\omega(\rho+V)}{3\Psi(2\omega+3)}$
\vspace{2mm}\\
3a) & $\rho+V \leq 0$ && $0 \leq \omega \leq \infty$ 
&& $\frac{4 \kappa^2 |\rho+V| \Psi}{2 \omega+3} \leq {\dot\Psi}^2 \leq \infty$
&& $0 \leq H^2 \leq \infty$
\vspace{2mm}\\
3b) & $\rho+V \leq 0$ && $-\frac{3}{2} \leq \omega \leq 0$ 
&& $\frac{4 \kappa^2 |\rho+V| \Psi}{2 \omega+3} \leq {\dot\Psi}^2 \leq \infty$ 
&& $\frac{2\kappa^2\omega(\rho+V)}{3\Psi(2\omega+3)} \leq H^2 \leq \infty$
\vspace{2mm}\\
4) & $\rho+V < 0$ && $-\infty < \omega < -\frac{3}{2}$ 
&& --
&& --
\vspace{4mm}\\
\hline
\end{tabular}
\end{center}
\caption{Constraints from the Friedmann equation on the values of 
$\dot{\Psi}\equiv y$ (\ref{y_allowed_inequality}) and 
$H$ (\ref{H_allowed_inequality}).}
\label{allowed_regions}
\end{table}

In terms of the phase space geometry the inequality 
(\ref{y_allowed_inequality}) is saturated on
a cylindrical 3-surface
\be
\mathcal{B}_H: B_H(x,y,H,\rho)\equiv \left( 2\omega(x)+3 \right) \frac{y^2}{12x^2} 
+ \frac{\kappa^2 (\rho + V(x))}{3x} =0 \, ,
\label{B_H}
\ee
which is parallel to the $H$ axis. 
In cases 2) and 3) it bounds the extent of the Friedmann
surface (\ref{Friedmann_surface}) in the $x, y, \rho$ directions,
i.e., when we project the Friedmann surface along the $H$ direction
to the ($x, y, \rho$) 3-plane, the projection lies within the bounds set by
$\mathcal{B}_H$. In case 1) the projection covers the entire 
($x, y, \rho$) 3-plane.
Correspondingly, 
the inequality (\ref{H_allowed_inequality}) is saturated on the 3-surface
\be
\mathcal{B}_{\dot{\Psi}}: B_{\dot{\Psi}}(x,y,H,\rho)\equiv 
\frac{x^2}{\omega^2(x)} \left( (2\omega(x)+3)H^2 - \frac{2 \kappa^2 \omega(x)}{3x} (\rho+V(x)) \right)=0 \, ,
\ee
which is parallel to the $y$ axis and bounds the extent of the Friedmann
surface in the $x, H, \rho$ directions.

In cases 2) and 3) the Friedmann 3-surface $\mathcal{F}$
tangentially touches
the cylindrical bounding 3-surface $\mathcal{B}_H$
along a 2-surface defined by the union 
$\mathcal{F}\cup\mathcal{B}_H$. 
The touching is tangential,
\be
 \nabla_i B_H \cdot T^i \Big|_{\mathcal{F}\cup\mathcal{B}_H} = 0 \,,
\ee
as the trajectories can not go through $\mathcal{B}_H$ to the
unphysical region. 
By substituting the condition (\ref{B_H}) into the
constraint equation (\ref{Friedmann_surface}) we see that
on the 2-surface $\mathcal{F}\cup\mathcal{B}_H$ 
the quantity $H+\frac{y}{2x}$ vanishes,
hence  $\mathcal{F}\cup\mathcal{B}_H$ joins
the above defined ``upper'' and ``lower'' regions 
of the Friedmann surface. 
We may ask, again relevant in cases 2) and 3) only,
whether the trajectories do 
cross between these two regions,
i.e. whether they do pass through  $\mathcal{F}\cup\mathcal{B}_H$. 
For this purpose let us introduce another 3-surface
\be
\bar{\mathcal{B}}_H: \bar{B}_H(x,y,H,\rho)\equiv H+\frac{y}{2x} =0 \, ,
\ee
distinct from $\mathcal{F}$ and $\mathcal{B}_H$, but with the property 
that its intersection with the Friedmann surface coincides with the union
of $\mathcal{F}$ and $\mathcal{B}_H$, i.e.,
$\mathcal{F}\cup \bar{\mathcal{B}}_H = \mathcal{F}\cup \mathcal{B}_H
= \mathcal{B}_H \cup \bar{\mathcal{B}}_H$.
From the scalar product
\beq
\label{crossing_regions}
 \nabla_i \bar{B}_H \cdot T^i \Big|_{\mathcal{F} \cup {\bar{\mathcal{B}}}_H} = 
\frac{\kappa^2}{2x} \left( (1-\rw)\rho + 2V \right) \,
\eeq
it follows that the 
trajectories pass through this intersection
from ``lower'' region where $H+\frac{y}{2x}<0$ to the ``upper'' region where 
$H+\frac{y}{2x}>0$ if 
$\left( (1-\rw)\rho + 2V \right)|_{\mathcal{F} \cup {\bar{\mathcal{B}}}_H}>0$ 
and in the reverse direction if 
$\left( (1-\rw)\rho + 2V \right)|_{\mathcal{F}\cup {\bar{\mathcal{B}}}_H}<0$.
Only if $\rw=1$ and $V|_{\mathcal{F}\cup {\bar{\mathcal{B}}}_H}=0$ 
do the trajectories, which are in the subspace 
$\mathcal{F}\cup {\bar{\mathcal{B}}}_H$, remain there in their entirety, 
and thus block the passage between the ``upper'' and ``lower'' regions.

In the next two sections we study the asymptotic behavior of solutions
by finding the fixed points and their properties.
Since the energy densities of different types of matter and the potential
$V(\Psi)$ scale differently under the change of the
scale factor of the Universe, 
one is justified to consider different regimes separately, specified by the
dominant component.

\section{Fixed points for potential domination ($V \not\equiv 0$, $\rho \equiv 0$)}

In the limit of vanishing matter density 
the phase space
shrinks to three dimensions 
$\{x \equiv \Psi, y \equiv \dot\Psi, H \}$, where the
Friedmann constraint restrains the physical trajectories to lie on a 
two-dimensional surface $\mathcal{F}$.
We may solve the Friedmann constraint for $H$, as (\ref{H_constraint}),
substitute it into Eq. (\ref{general_dynsys_y}), and thus in effect reduce
the system 2-dimensional:
\beq \label{V_dynsys}
\left \{ \begin{array}{rcl}
\dot{x} & = & y  \\
\dot{y} & = & 
\left( \frac{3}{2x} - \frac{1}{2 \omega(x) + 3}\, \frac{d \omega}{dx}
\right) \, y^2
\mpp  \frac{1}{2x} \sqrt{3 (2 \omega(x) + 3) y^2 + 12 \kappa^2 x V(x)}\ y 
+ \frac{2 \kappa^2}{2 
\omega(x) + 3} \, 
\left(2 V(x) - x \, \frac{dV}{dx} \right)
\,.
\end{array}
\right.
\eeq
This constitutes a projection of the trajectories on 
the original two-dimensional constraint 
surface in $(x, y, H)$ to the $(x,y)$ plane. 
The projection yields two ``sheets'': the ``upper sheet'' 
marked by the $\mxp$ sign, and
the ``lower sheet'' 
marked by the the $\xpp$ sign in Eq. (\ref{V_dynsys}).
In three dimensions the former corresponds to the ``upper'' region 
where $H+\frac{y}{2x} >0$,
while the latter to the ``lower'' region 
where $H+\frac{y}{2x} <0$.

The consideration of only the variables $\{x,y\}$
and Eqs. (\ref{V_dynsys}) is in principle sufficient
to follow the dynamics
as the value of $H$ can at each point be computed from the constraint 
(\ref{H_constraint}). 
The only complication arises in cases 2) and 3) along the boundary 
$y^2 = \frac{4 \kappa^2 x |V(x)|}{|2\omega(x)+3|}$ which marks the extent 
of the $(x,y)$-plane accessible for physical trajectories. 
In three dimensions this boundary corresponds to the border 
$\mathcal{F}\cup\mathcal{B}_H$ where $H+\frac{y}{2x}=0$ and the
``upper'' and ``lower'' regions meet. 
From the discussion
around Eq. (\ref{crossing_regions}) we know that with the exception of 
some special cases the trajectories do traverse from one region to another,
and thus in the two-dimensional projection picture must
change the ``sheet''. 
However, along the boundary the Friedmann surface is positioned
perpendicularly to the $(x,y)$-plane and the projection (\ref{V_dynsys})
is then unable to
encode the full dynamics of the system.
Therefore with Eq. (\ref{V_dynsys}) it is possible to follow
a trajectory on one ``sheet''
forwards until it reaches the boundary or 
backwards until it starts from the boundary,
while the step of changing the sheet remains to be accounted
by relying on the full system 
(\ref{general_dynsys_x})-(\ref{general_dynsys_rho}).
Still, the system (\ref{V_dynsys}) is adequate for finding and 
describing the fixed points, 
at least as long as the prospective fixed points do not 
reside on the border of the two ``sheets''.

Standard procedure reveals that 
the dynamical system (\ref{V_dynsys}) is endowed with two fixed points, 
Table \ref{V_fp_eigenvalues} lists their conditions and eigenvalues. 
The first fixed point $\Psi_{\bullet}$ satisfies 
\be
\label{V_Psi_bullet}
\frac{\dd V}{\dd \Psi}\Big|_{\Psi_{\bullet}} \Psi_{\bullet} - 2 V(\Psi_{\bullet})=0 \, ,
\ee
which matches the second limit (\ref{second_GR_limit}),
discussed in Sec. 2.
It is rather 
surprising to note that a local extremum of the potential,
$\frac{\dd V}{\dd \Psi}=0$,
does provide
 a fixed point only if the value of the potential vanishes at this point,
$V(\Psi_{\bullet})=0$.
In the latter case the eigenvalues tell that  
if this extemum is a maximum, 
$\frac{d^2 V}{d\Psi^2}<0$, the nature of this point is saddle, 
while a local minimum, $\frac{d^2 V}{d\Psi^2}>0$, is non-hyperbolic.
But generally local minima with non-vanishing value of the potential
do not figure as fixed points, while at the same time it is possible
to have a fixed point residing 
on an arbitrary steep slope of the potential. 
This puzzling feature, however, becomes more meaningful when translated
into the Einstein
frame. As explained in Sec. 6 the condition (\ref{V_Psi_bullet}) gives
a local 
extremum of the Einstein frame potential.
\begin{table}[t] 
\begin{center}
\begin{tabular}{lclcl}
 & \quad & Condition & \quad & Eigenvalues
\\ \hline \\
$\Psi_{\bullet} $ && 
$\frac{dV}{d\Psi}|_{\Psi_{\bullet}} \Psi_{\bullet} - 2 V(\Psi_{\bullet})=0$ &&
$\pmp \left[ -\sqrt{ \frac{3 \kappa^2 V}{4 \Psi}} \pm 
\sqrt{ \frac{\kappa^2}{2\omega+3} \left( (6\omega + 25) \frac{V}{4\Psi} - 2 \Psi \frac{d^2V}{d\Psi^2} \right) } \right]_{\Psi_{\bullet}}$  \vspace{2mm}\\
$\Psi_{\star} $ && 
$\frac{1}{2\omega(\Psi_{\star}) +3} = 0$, 
$\frac{1}{(2\omega(\Psi_{\star})+3)^2} \frac{\dd \omega}{\dd \Psi} \neq 0$ &&
$\pmp \left[ -\sqrt{ \frac{3 \kappa^2 V}{4 \Psi} } \pm 
\sqrt{ \frac{4 \kappa^2}{(2\omega +3)^2} \frac{d\omega}{d\Psi} \left(\Psi \frac{dV}{d\Psi} - 2V \right) + \frac{3 \kappa^2 V}{4 \Psi} } \right]_{\Psi_{\star}}$  
\vspace{2mm}\\
\hline
\end{tabular}
\end{center}
\caption{Fixed points and their eigenvalues for the $V \not\equiv 0$, $\rho \equiv 0$ case. 
}
\label{V_fp_eigenvalues}
\end{table}

The second fixed point $\Psi_{\star}$ satisfies
\be
\label{V_Psi_star}
\frac{1}{2\omega(\Psi_{\star}) +3} = 0 \,,
\qquad
\frac{1}{(2\omega(\Psi_{\star})+3)^2}\frac{\dd \omega}{\dd \Psi}\Big|_{\Psi=\Psi_{\star}} \neq 0,
\ee
i.e. exactly the same conditions (a)-(d)
as the limit of general relativity for flat
FLRW STT cosmology, discussed in the end of Sec. 2.
Again, the eigenvalues are listed in Table \ref{V_fp_eigenvalues}. 
In particular we see
that on the ``upper'' sheet 
this point is an attractor if $V>0$ and
$\frac{4 \kappa^2}{(2\omega +3)^2} \frac{d\omega}{d\Psi} \left(\Psi \frac{dV}{d\Psi} - 2V \right) < 0$, while on the ``lower'' sheet it 
can not be an attractor at all.
(In principle it is also 
possible that the points $\Psi_{\bullet}$ and $\Psi_{\star}$
coincide. Yet, 
the properties of such a combined point are difficult to determine 
without knowing the exact form of $\omega$ and $V$.)

From Eq. (\ref{H_constraint}) it is straightforward to compute that 
the values of $H$
corresponding to the fixed points $\Psi_{\bullet}$ and $\Psi_{\star}$ are
$H_{\bullet}= \pmp \sqrt{ \frac{\kappa^2 V(\Psi_{\bullet})}{3 \Psi_{\bullet}}}$ and $
H_{\star}= \pmp \sqrt{ \frac{\kappa^2 V(\Psi_{\star})}{3 \Psi_{\star}}}$,
respectively. The result, which mimics 
de Sitter evolution in general relativity, was expected, since we 
saw in Sec. 2 that under the fixed point conditions (\ref{V_Psi_bullet})
and (\ref{V_Psi_star}) the full
STT equations (\ref{general_dynsys_x})-(\ref{general_dynsys_H}) reduce to the equations of 
general relativity.

So, having a model of STT with given $\omega(\Psi)$ and $V(\Psi)$,
one has to solve the conditions (\ref{V_Psi_bullet}) and (\ref{V_Psi_star})
to find the values of $\Psi$ where fixed points occur. 
To determine the nature of these points it is necessary to 
compute the eigenvalues at these points, 
possibly using an appropriate limiting procedure.
For the benefit of the reader, may we recall that when both eigenvalues are 
real and negative then we have a stable node (attractor), 
while real and positive eigenvalues indicate an unstable node (repeller).
Complex eigenvalues with a negative real part 
give a stable focus (spiralling attractor), 
while a positive real part indicates an unstable focus. 
One positive and one negative real eigenvalue occur with a saddle point,
but if the real part vanishes, the point is classified as non-hyperbolic
and linear stability analysis is not enough to determine the behaviour of
solutions around it.

As an illustration let us consider an example
\be
\label{example omega V}
\omega(\Psi)=\frac{3 \Psi}{2(1-\Psi)} \,, \qquad
V(\Psi) = -2 (\Psi - 0.2 )^3 + 3 (\Psi - 0.2)^2 \,,
\ee
chosen to demonstrate some typical features that may crop up in a
generic scalar-tensor theory. 
The shape of $\omega$ and $V$, along with 
the phase portraits 
are shown on Fig. 1. 
To briefly go through the main features let us first recall that 
while the phase space is 3-dimensional $(\Psi, \dot{\Psi}, H)$ the
physical trajectories lie on the 2-dimensional Friedmann surface 
which is comprised of two sheets.
The domain $\Psi \in (0, 1]$ belongs to case 1a) of Table \ref{allowed_regions}, meaning
the values of $\dot{\Psi}$ are not limited, and the 
``upper'' and ``lower'' sheets are separate. 
On the other hand, 
the domain  $\Psi \in [1, 1.7]$ belongs to the case 2), where 
the values of $\dot{\Psi}$ are limited and the 
``upper'' and ``lower'' sheet meet along the $H+\frac{\dot{\Psi}}{2\Psi}=0$ 
line. 
Finally, the domain  $\Psi \in (1.7, \infty)$ falls under the case 4) 
where the Friedmann constraint does not have any real solutions.

\begin{figure}[tp]
\phantom.
\hspace{-2cm}
\includegraphics[width=9.7cm,height=9cm]{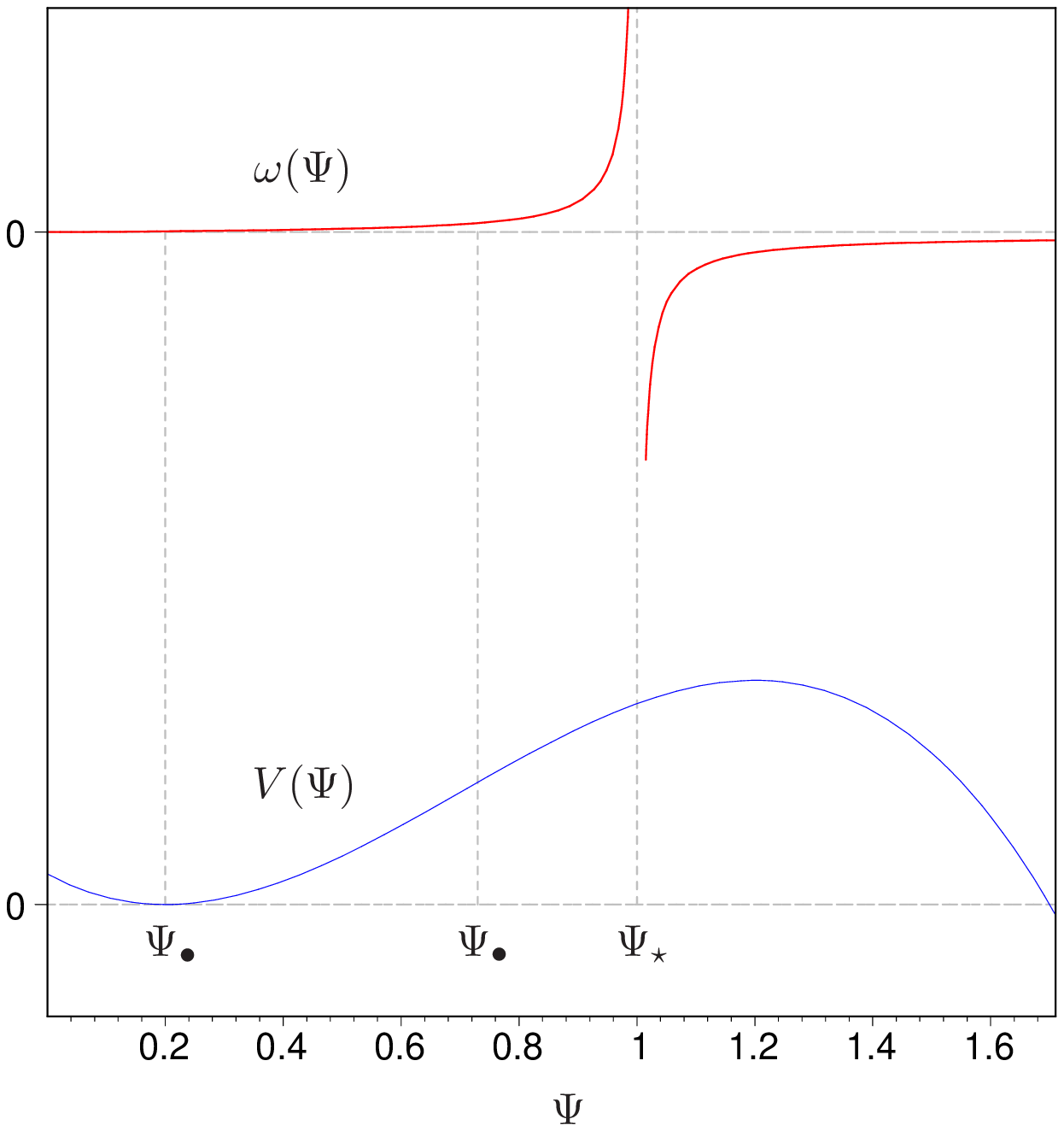}
\hspace{-1.5cm}
\includegraphics[width=10cm,height=10cm]{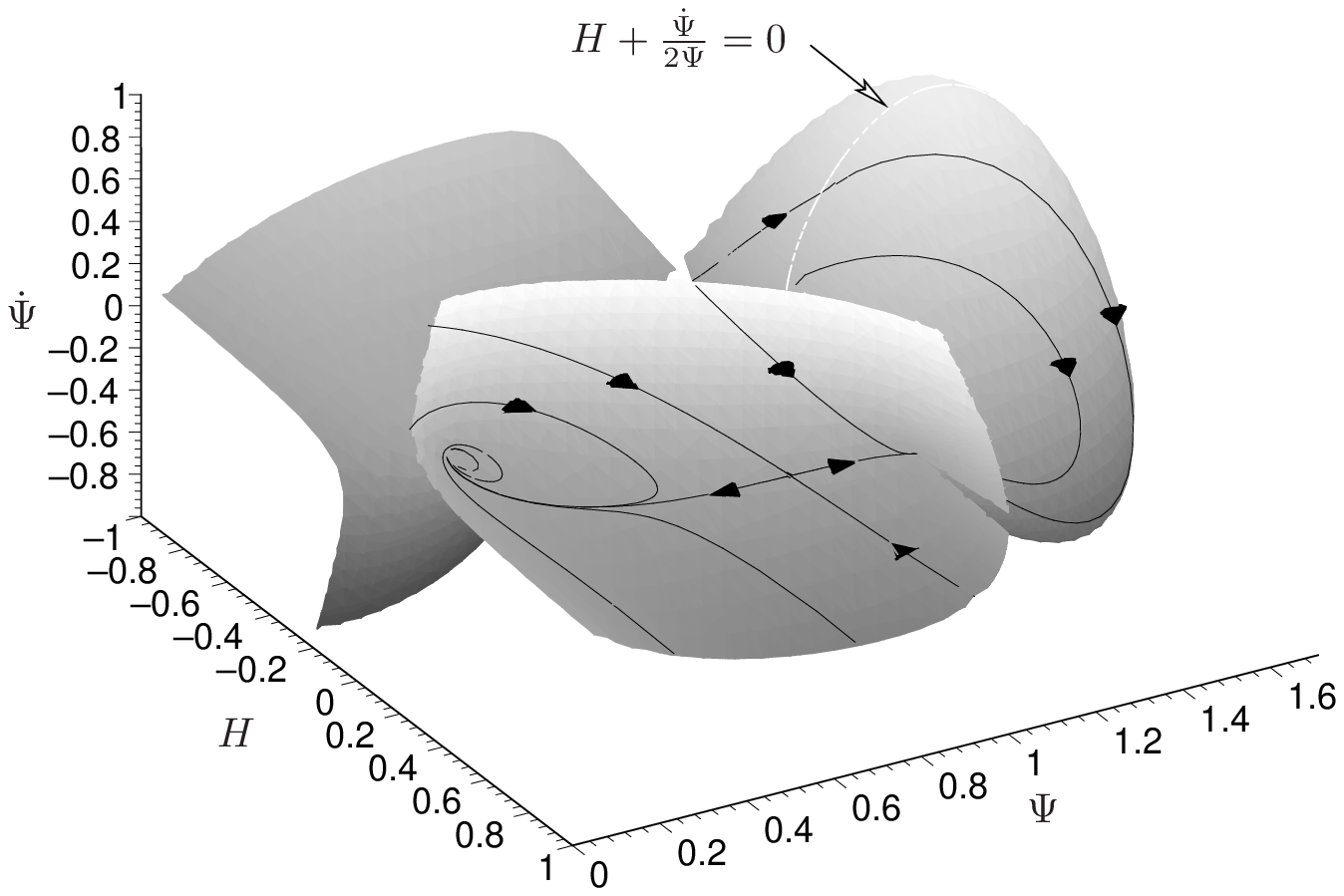}
\vspace{1cm}
\linebreak
\phantom.
\vspace{0cm}
\phantom.
\hspace{-2.5cm}
\includegraphics[width=9cm,height=9cm]{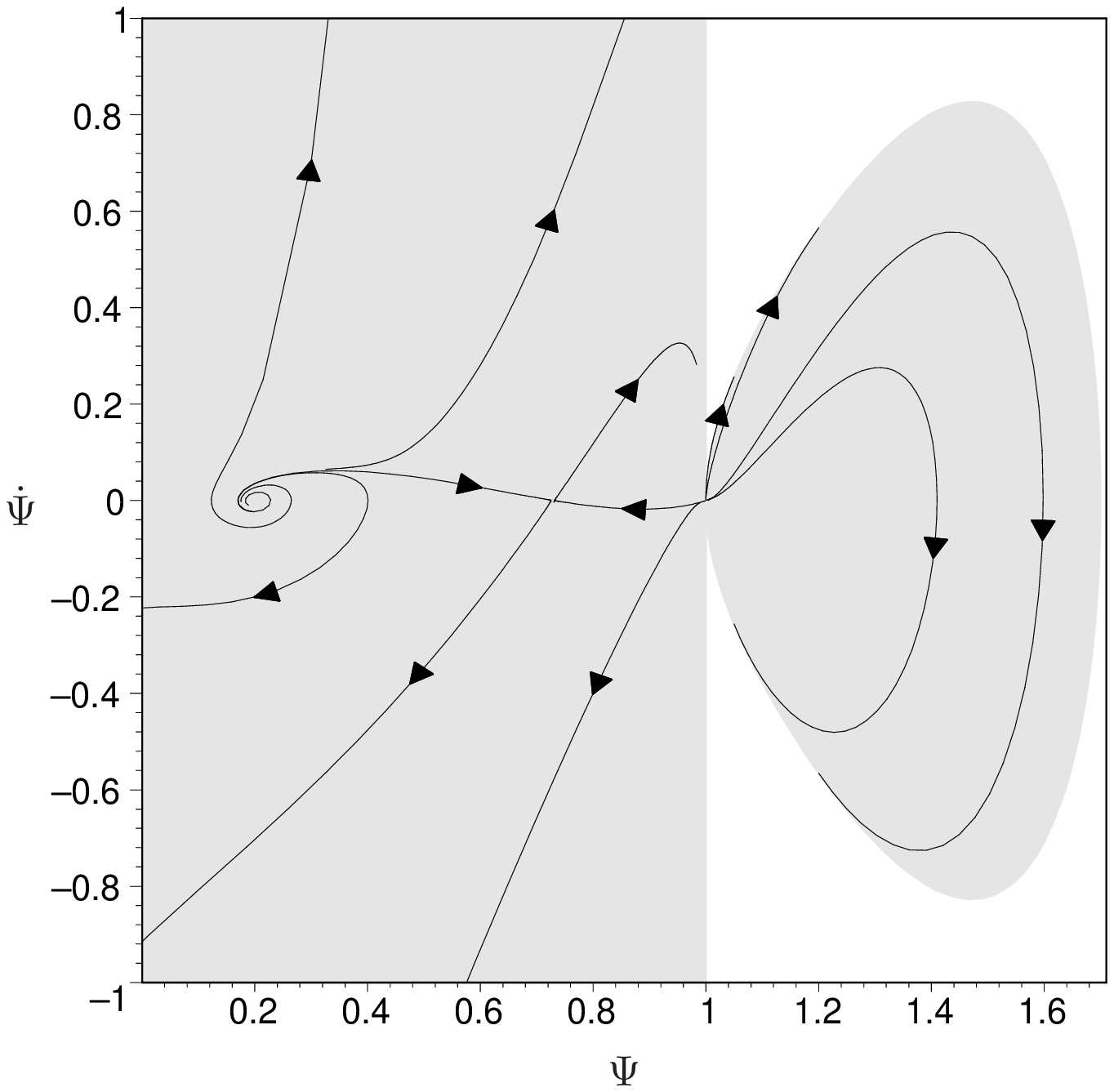}
\hspace{-0.5cm}
\includegraphics[width=9cm,height=9cm]{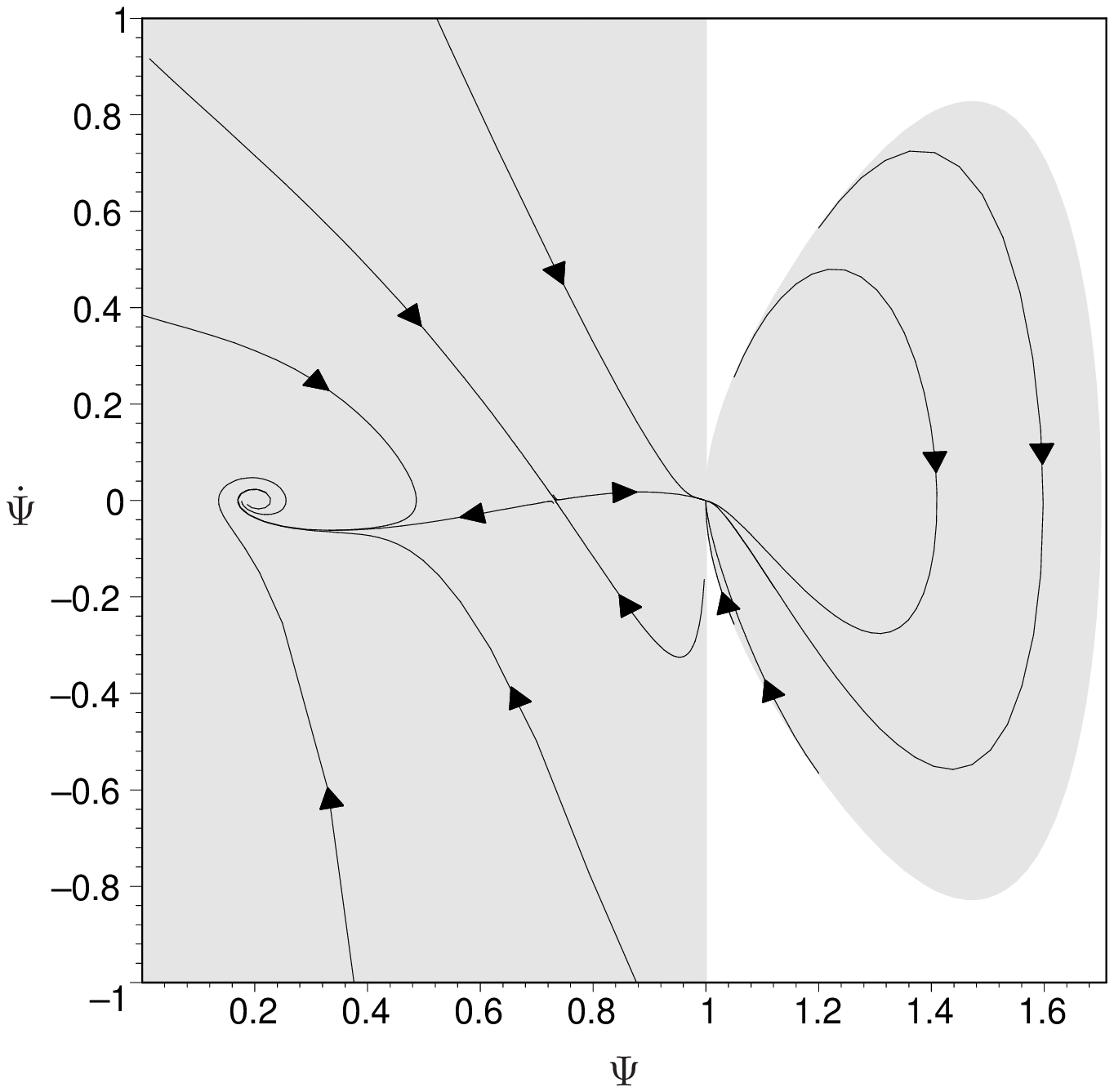}
\caption{Illustration to the example (\ref{example omega V}) described 
in the text: the shape of 
$\omega$ and $V$ (up, left); 3-dimensional phase space with two sheets 
where the physical trajectories reside (up, right); 2-dimensional projection
of the ``lower'' sheet (down, left), and ``upper'' sheet (down, right).}
\label{plots}
\end{figure}

There are three fixed points coming in two sets, one set for
the ``upper'' and another for the ``lower'' sheet. 
Two of the fixed points 
satisfy the
condition (\ref{V_Psi_bullet}) and happen to 
have the same corresponding nature on both sheets: 
the point at $\Psi_{\bullet}=0.2$ (where $V=0$) is non-hyperbolic,
while the other point at $\Psi_{\bullet}=0.73$ is a saddle. 
The third fixed point $\Psi_{\star}=1$ satisfies (\ref{V_Psi_star})
and is an attractor on the 
``upper'' sheet, but a repeller on the ``lower'' sheet.
Fig. 1 (up, right) illustrates how the trajectories in the $\Psi \in [1, 1.7]$ domain
start at the ``lower'' sheet repeller $\Psi_{\star}$, cross over to the 
``upper'' sheet, and end up at the ``upper'' sheet attractor $\Psi_{\star}$.
This behavior is reflected on the projection, which shows the
``lower'' sheet trajectories (Fig. 1, down, left) 
running to the boundary in order to 
just emerge on the ``upper'' sheet boundary 
(down, right).

\section{Fixed points for matter domination
($V \equiv 0$, $\rho \not\equiv 0$)}

In the case of cosmological matter ($\rho>0$) and vanishing scalar potential
the Friedmann constraint restricts the solutions onto 
a three-dimensional surface in four phase space dimensions 
$(x \equiv \Psi, y \equiv \dot\Psi, H, \rho)$.
However, the system is amenable to a change of the time variable, first 
used by Damour and Nordtvedt \cite{dn} in the Einstein frame, 
that allows to combine the field equations into 
a dynamical equation for 
the scalar field which does not manifestly contain the scale factor or matter 
density. 
In the Jordan frame this amounts to defining a new time variable $p$
\cite{san}
\be \label{jt_ptime}
dp = h_c dt \equiv \left| H + \frac{\dot{\Psi}}{2\Psi} \right| dt \,.
\ee
Then from Eqs. (\ref{00})--(\ref{deq})
the following ``master'' equation for the scalar field can be 
derived \cite{san, meie2}:
\beq
\Psi'' 
\mpp {\frac {(2 \omega(\Psi) +3) (1-\rw) }{ 8 \Psi^2}} \, \Psi'^3 
- \left( \frac{3 (1+\rw)}{4 \Psi} 
- \frac{1}{2 \omega(\Psi) +3} \frac{d\omega(\Psi)}{d\Psi} \right) \, \Psi'^2 &&
\nonumber \\
\pmp \frac{3 (1 - \rw) }{2}\, \Psi' 
- \frac{3(1-3 \rw)}{(2\omega(\Psi)+3)}\, \Psi &=& 0 \,,
\label{j_master}
\eeq
where primes denote the derivatives with respect to $p$. 
The transformation of the time coordinate (\ref{jt_ptime}) 
managed to align the phase space trajectories in a way that made 
a projection into two dimensions possible. 
Like in the previous section 
the upper signs in Eq. (\ref{j_master}) correspond to 
the ``upper'' sheet where the quantity 
$H + \frac{\dot{\Psi}}{2\Psi}$ is positive, 
while the lower signs to the ``lower'' sheet where the quantity 
$ H + \frac{\dot{\Psi}}{2\Psi}$ is negative. 

The ``master'' equation can be relied on as long as
$h_c= |H + \frac{\dot{\Psi}}{2\Psi}|$ is finite.
At $\Psi=0$ the quantity $h_c$ diverges making the $t$-time to stop 
with respect to the $p$-time. Hence all $t$-time trajectories 
with finite $\dot{\Psi}$  get mapped to $\Psi'=0$, giving 
a false impression of a fixed point there. 
However, in Sec. 2 we concluded that  $\Psi=0$ comes with a 
space-time singularity and exclude it from present analysis.

The other problematic points can be discussed by noticing that 
in terms of the new time variable $p$ 
the Friedmann constraint (\ref{00}) can be written as 
\beq \label{jf_ptime_friedmann}
h_c^2= \frac{\kappa^2  \, \rho 
}{3\Psi \left( 1- 
\frac{(2\omega(\Psi)+3)\, \Psi'^2}{12 \Psi^2} \right)} \,.
\eeq
To keep $h_c$ real, the right hand side
of Eq. (\ref{jf_ptime_friedmann})
must be nonnegative, thus constraining the dynamically allowed regions of the
two-dimensional phase space ($\Psi, \Psi'$) of the scalar field 
(see Table \ref{matter_allowed_regions}).
\begin{table}[t] 
\begin{center}
\begin{tabular}{llclccc}
&  & \qquad &  & \quad & Allowed range & Allowed range
\\
&  & \qquad &  & \quad & of $\dot\Psi$ 
& of $\Psi'$ 
\vspace{2mm}\\
\hline \\
1) &  $\rho > 0$  && $2\omega +3 > 0$ && $0 \leq {\dot\Psi}^2 < \infty$ 
& $0 \leq {\Psi'}^2 < \frac{12\Psi^2}{2\omega+3}$ 
\vspace{2mm}\\
2) & $\rho > 0$ && $2\omega +3 \leq 0$ && $0 \leq {\dot\Psi}^2 \leq 
\frac{4 \kappa^2 \rho \Psi}{|2 \omega+3|}$ 
&  $0 \leq {\Psi'}^2 < \infty$
\vspace{4mm}\\
\hline
\end{tabular}
\end{center}
\caption{Constraints from the Friedmann equation on the scalar field phase 
space in $t$-time from Eq. (\ref{y_allowed_inequality}) and 
$p$-time from Eq. (\ref{jf_ptime_friedmann}).}
\label{matter_allowed_regions}
\end{table}
At the boundaries of the allowed regions the time transformation 
(\ref{jt_ptime}) fails to be meaningful.
In case 1) the boundary
\beq \label{rho Friedmann bound}
\Psi'^2 = \frac{12\Psi^2}{2\omega(\Psi)+3} \,
\eeq
is characterized by $t$-time stopping with respect to $p$-time, and
as a result finite $\Psi'$ corresponds to 
${\dot \Psi}^2 \rightarrow \infty$. Considering $t$-time to be physical
permits us to exclude this boundary from analysis on physical grounds.
In case 2) the limit $\Psi' \rightarrow \infty$ corresponds
to finite $\dot{\Psi}$, since $p$-time stops with respect to $t$-time.
As $h_c = |H + \frac{\dot{\Psi}}{2\Psi}|=0$ 
the latter boundary also marks the joint of the
``upper'' and ``lower'' sheets. 
Therefore, 
in $p$-time it is not possible to follow the passage of trajectories 
from one sheet to another, but as the general considerations presented after
Eq. (\ref{crossing_regions}) inform us, the trajectories do
pass from the ``lower'' to ``upper'' sheet if $\rw<1$ and vice versa 
if $\rw>1$, while in the $\rw=1$ case the passage is blocked by trajectories 
which lie entirely on this boundary.
Despite the shortcoming that the two-dimensional 
``master'' equation (\ref{j_master}) is not able to capture
all the details of the full dynamics in the four-dimensional phase space,
we can still use it for finding the fixed points for $\Psi$ 
as long as the fixed 
points do not reside on the problematic boundary.

Let us consider the cosmological matter behaving like dust ($\rw=0$) first.
Introducing $\Psi=x$, $\Psi'=z$ enables to write 
Eq. (\ref{j_master}) as a dynamical system: 
\beq
\left \{ \begin{array}{rcl}
x' & = & z  \\
z' & = & 
\pmp \frac {2 \omega(x) +3}{8 x^2 } \, z^3 
+ \frac{6\omega(x)+9-4x \frac{d\omega(x)}{dx}}{4x (2 \omega(x) +3)} \, z^2 
\mpp \frac{3}{2}\, z 
+ \frac{3x}{2\omega(x)+3}
\,.
\end{array}
\right.
\eeq
An argument completely analogous to the one put forth for Eq. 
(\ref{taylor_expansion}), reveals a single fixed point, 
satisfying the conditions (a)-(d)
dubbed as the limit of general relativity for flat FLRW STT in Sec. 2.
The corresponding eigenvalues, to be evaluated at the fixed point coordinate,
are given in Table \ref{matter_fp_eigenvalues}.
In particular, this point is an attractor on the ``upper'' sheet if
$\frac{\dd \omega}{\dd \Psi} >0$, while on the ``lower'' sheet attractor 
behavior is not possible.

To complete the analysis it is important to verify that the fixed point 
just found in the $p$-time does indeed
correspond to a fixed point in the cosmological $t$-time. 
But since we have excluded the boundary (\ref{rho Friedmann bound})
and consider only the trajectories with finite $h_c$, it
is immediate that $\Psi'=0$ implies $\dot{\Psi}=0$, due to (\ref{jt_ptime}).
From Eq. (\ref{H_constraint}) now it also follows
that at the fixed point the evolution of the universe obeys the usual
Friedmann equation from general relativity,
$H_{\star}^2=\frac{\kappa^2 \rho}{3 \Psi_{\star}}$.
This is expected, as the fixed point conditions were identical to the
general relativity limit.

\begin{table}[t] 
\begin{center}
\begin{tabular}{lclclcl}
Case & \qquad & Fixed point & \qquad & Condition & \quad & Eigenvalues \vspace{2mm}\\
\hline \\
$\rw=0$&&
$\Psi_{\star}$ && $\frac{1}{2 \omega(\Psi_{\star})+3}=0$, 
$\frac{1}{(2\omega(\Psi_{\star})+3)^2} \frac{\dd \omega}{\dd \Psi} \neq 0$ && 
$\pmp \left[ -\frac{3}{4} \pm \frac{3}{4} 
\sqrt{1 - \frac{32}{3} \frac{\Psi}{(2 \omega+3)^2} \frac{d \omega}{d \Psi}}
\right]_{\Psi_{\star}}$ \vspace{2mm}\\
$\rw={1 \over 3}$ && none &&
\vspace{4mm}\\
\hline
\end{tabular}
\end{center}
\caption{Fixed points and their eigenvalues for the $\rho \not\equiv 0$,
$V \equiv 0$ case.}
\label{matter_fp_eigenvalues}
\end{table}

The dynamical system in the radiation dominated regime ($\rw=\frac{1}{3}
$) reads 
\beq \label{kiirgus}
\left \{ \begin{array}{rcl}
x' & = & z  \\
z' & = & 
\pmp \frac {2 \omega(x) +3}{12 x^2 } \, z^3 
+ \frac{8\omega(x)+12-4x \frac{d\omega(x)}{dx}}{4x (2 \omega(x) +3)} \, z^2 
\mpp z 
\,.
\end{array}
\right.
\eeq
There are no fixed points. For small values of $\Psi'=z$ 
the system is ruled by friction on the ``upper'' sheet, as the $\mxp$ sign 
of the dominating term forces the vector flow to converge
to the $z' = 0$ axis.
On the ``lower'' sheet, the the effect is the opposite (anti-friction).

\section{Discussion}

%
A lot of work in FLRW scalar-tensor cosmology has been performed in the Einstein 
frame which is obtained from the Jordan frame by two transformations 
\cite{dicke, flanagan}:
(1) a conformal transformation of the
metric ${\tilde{g}_{\mu \nu}^{(E)}} = \Psi g_{\mu \nu}^{(J)}$, 
followed by a coordinate transformation to keep the FLRW form of the
line element, $d\tilde{t} = \sqrt{\Psi} dt$,
and (2) a
redefinition of the scalar field 
\be \label{sf_transformation}
(2 \omega (\Psi) +3) (d\Psi)^2 =
4 \Psi^2 (d \phi)^2. 
\ee
The two frames are mathematically equivalent and hence also physically equivalent, 
as long as these two transformations are regular \cite{dicke, weinberg, faraoni, flanagan, faraoni2}. 
Both transformations become singular in the limit $\Psi \rightarrow 0$, while the latter 
transformation is also singular in the limit $2\omega(\Psi)+3 \rightarrow 0$. 
These singularities were scrutinized in Refs. \cite{abramo_2003, faraoni_2004, faraoni_2008}
with the conclusion that both frames retain equivalence in the sense that in both frames the 
Cauchy problem fails to be well posed in these limits. In Sec. 2 we also observed in passing 
that approaching these two limits generically leads to a space-time singularity in the Jordan frame.

The transformation (\ref{sf_transformation}) happens to be singular in the limit of general relativity,
$\frac{1}{2\omega(\Psi)+3} \rightarrow 0$, as well. 
Ref. \cite{meie3} also gives explicit examples of coupling functions $\omega(\Psi)$ where 
the phase portraits for regions containing the GR limit are qualitatively inequivalent in the two frames. 
Therefore it is not guaranteed that the properties of the $\Psi_{\star}$ fixed point which resides 
at the general relativity limit are exactly the same in the Einstein frame. 
To actually compare the two frames in this respect,
it would be necessary to perform a similar fixed point analysis in the Einstein frame.

The other fixed point, $\Psi_{\bullet}$, however, should have the same properties in the 
Einstein frame as in the Jordan frame, as long as it resides in the region where the 
transformations between the two frames are regular.

The conformal transformation relates the Hubble parameters in the two frames as
\be
\label{H_E H_J}
\tilde{H}_E = \frac{1}{\sqrt{\Psi}}\left( H_J +\frac{\dot\Psi}{2\Psi}\right) \, .
\ee
Therefore the sign of the quantity $H_J +\frac{\dot\Psi}{2\Psi}$ 
which in our Jordan frame analysis distinguished the ``upper'' and ``lower'' 
sheet of the Friedmann surface, has a clear interpretation in the Einstein
frame as indicating the expanding or collapsing universe.
Also, as the scalar potentials in the two frames are related by
\be
\label{V_E V_J}
V_E (\phi(\Psi)) = \frac{1}{\Psi^2}V_J(\Psi) \,,
\ee
it follows that the fixed point $\Psi_{\bullet}$ given by Eq. 
(\ref{V_Psi_bullet}) indeed corresponds to the local extremum 
of $V_E$ \cite{faraoni3}:
\be
\frac{\dd V_E}{\dd \phi}\Big|_{\phi(\Psi_{\bullet})} = 
\frac{1}{\Psi_{\bullet}^3}
\left[ \frac{\dd V}{\dd \Psi} \Psi - 2 V \right]_{\Psi_{\bullet}} 
\frac{\dd \Psi}{\dd \phi}\Big|_{\Psi_{\bullet}} = 0 \,.
\ee

Among the two types of fixed points we found,
$\Psi_{\bullet}$ 
and $\Psi_{\star}$,
it is the latter which deserves particular interest, since 
it satisfies the PPN limit of general relativity and
good conformity with Solar System experiments is guaranteed.
The existence of $\Psi_{\star}$ relies on the conditions
(a)-(d) given in Sec. 2, while its nature (attractor or other)
is determined by the eigenvalues listed in Tables \ref{V_fp_eigenvalues} 
and \ref{matter_fp_eigenvalues}.
Previous phase space analyses
performed for specific examples of coupling functions and potentials 
which have considered this point \cite{meie2, meie3}
are in accord with our general results.
Still, we may also attempt a comparison with relevant studies 
carried out by other methods.

Damour and Nordtvedt \cite{dn} 
argue that for 
large classes  of coupling functions $\alpha^2 (\phi) = 
\frac{1}{2\omega (\Psi) +3}$  solutions are driven to the value of scalar
field where the coupling function $\alpha$ vanishes, 
the scalar field decouples and 
we are left with general relativity. Their investigations were performed
in the Einstein frame
using approximate late time solutions and their results are in qualitative
agreement with our results: for radiative matter, there is no
specific fixed point, for dust matter general relativity is
a fixed point.
For instance taking linear coupling, 
$\alpha (\phi) = K \phi$ with $K = const>0$, in the matter dominated regime,
they found that for $K<\frac{3}{8}$ the system exhibits damped behavior,
while for $K>\frac{3}{8}$ the behavior is damped-oscillatory.
In our analysis this can be compared to the $\Psi_\star$ fixed point. 
Plugging in the coupling to the eigenvalues in Table \ref{matter_fp_eigenvalues} gives
$-\frac{3}{4} \pm \frac{3}{4}( \sqrt{1 - \frac{16}{3} K})$, 
indicating a stable node (attractor) for $K < \frac{3}{16}$ and
stable focus (spiralling attractor) for $K > \frac{3}{16}$.
The slight difference in the numerical factor may reflect the difference between 
the frames, or perhaps can be attributed
to the fact that in making the approximation 
Damour and Nordtvedt drop some terms in the scalar field equation.

Recently Barrow and Shaw \cite{bs} analysed the asymptotic behavior 
of homogeneous and isotropic cosmological solutions of  
general scalar-tensor theory 
containing a two-component perfect fluid: 
vacuum 'dark energy' ($p_{0} = - \rho_{0}$, essentially a constant potential)
and sub-dominant matter density.
They
demonstrate that if STT cosmology 
evolves toward the de Sitter 
limit at late time, then there exists 
an asymptotic value of the scalar field $\Psi_{\infty} \in (0, \infty)$  such 
that for  $\Psi \rightarrow \Psi_{\infty}$ it holds that 
$\omega \rightarrow \infty$ and  
$\frac{1}{\omega^{2 + \epsilon}} \frac{d\omega}{d\Psi} 
\rightarrow 0$ for any $\epsilon >0$. 
In our analysis of Sec.~4 this can be compared to
the solutions closing in on the fixed point $\Psi_{\star}$ 
yielding de Sitter expansion.
The conditions for the existence of $\Psi_{\star}$, 
(a) $\frac{1}{2\omega+3} \rightarrow 0$ and (c)  
$\frac{1}{(2 \omega +3)^2}\frac{d\omega}{d\Psi}\vert_{_{\Psi=\Psi_{\star}}}
 \neq 0$ are slightly stronger than the ones of Barrow and Shaw. 
Namely, if  $\Psi_{\star}$ is identified with $\Psi_{\infty}$,
then (a) and (c) imply 
the conditions given by Barrow and Shaw,
but the reverse does not neccessarily follow, 
since (c) may not be satisfied.

Finally let us note that scalar-tensor theory has some features 
which allow it to be seen as a simplistic
toy model of 
the low energy effective actions 
derived from string/M-theory compactifications. 
Namely, the latter usually involve a number of scalar fields 
(moduli of the compactification) and some of these are
coupled to the Ricci scalar in the Jordan/string frame description.
For phenomenological reasons it is important to have the moduli stabilized 
at fixed values.
Typical scenarios of string theory moduli stabilization involve
generating a scalar potential whereby the moduli can stabilize at
the minimum of the potential \cite{string_moduli}. 
In our notation this parallels to invoking the fixed point $\Psi_{\bullet}$.
However, it would be interesting to see whether the other fixed point, 
$\Psi_{\star}$, occuring at the 
singularity of the scalar field kinetic term, can also be generalized for 
string theory compactifications and
what role can it play in stabilizing the string theory moduli.

\section{Conclusion}

We have considered flat FLRW cosmological models in general 
scalar-tensor theories with arbitrary coupling function $\omega(\Psi)$
and scalar potential $V(\Psi)$ in the Jordan frame. Using the methods
of dynamical systems we have described the general 
geometry of the phase space and
found the scalar field 
fixed points in two distinct 
asymptotic regimes: potential domination 
($V\not\equiv 0, \rho \equiv 0$), and
matter domination ($V \equiv 0, \rho \not\equiv 0 $). 
In nutshell there are two types of fixed points arising
from different mechanisms: $\Psi_{\bullet}$ from a condition on
the potential (equalling the local extremum
of the Einstein frame potential) 
and $\Psi_{\star}$ from the singularity
of the scalar field kinetic term.
Approaching both types of fixed points 
the cosmological equations coincide
with those of general relativity, yielding de Sitter expansion in
the potential domination case and Friedmann evolution in the matter domination
case.
However, for the Solar System experiments in the PPN framework
 only the fixed points of $\Psi_{\star}$ type give
predictions identical with those of general relativity.
The nature of fixed points (attractor or otherwise) depends
on the functional forms of
$\omega(\Psi)$ and $V(\Psi)$ according to corresponding 
eigenvalues given in Tables \ref{V_fp_eigenvalues} and \ref{matter_fp_eigenvalues}. 
Therefore, in Jordan frame analysis, general relativity is an attractor
for a large class of scalar-tensor models, but not for all.

Provided the transformation relating the Jordan and Einstein frame is regular, 
there is an exact correspondence between the two frames and the Jordan frame 
phase space results should carry over to the Einstein frame. 
This is the case for the fixed point $\Psi_{\bullet}$.
However, as the transformation of the scalar field fails to be regular in the limit of 
general relativity, the properties of the $\Psi_{\star}$ fixed point may be altered 
in the Einstein frame. 
To establish whether or how the correspondence holds in this case calls for a separate 
matching investigation in the Einstein frame.

\bigskip
{\bf Acknowledgements}

We are grateful to the anonymous referee for clarifying questions.
This work was supported by the Estonian Science Foundation
Grant No. 7185 and by Estonian Ministry for Education and Science
Support Grant No. SF0180013s07. 
MS also acknowledges the 
Estonian Science Foundation Post-doctoral research grant No. JD131.

\end{document}